\documentclass[12pt]{article}
\usepackage{amsfonts}
\usepackage{amsmath}
\usepackage{amssymb}
\usepackage{graphicx}
\usepackage{color}
\usepackage[all, knot]{xy}
\usepackage{tikz}
\usepackage{array}

\usepackage{ulem}

\usepackage[utf8]{inputenc}
\usepackage{epstopdf}
\usepackage[footnotesize]{caption}
\usepackage{amsthm}
\usepackage{enumitem}
\usepackage{mathrsfs}

\usepackage[margin=3cm]{geometry}

\def \be {\begin{equation}}
\def \ee {\end{equation}}
\def \bea {\begin{eqnarray}}
\def \eea {\end{eqnarray}}
\def \nn {\nonumber}

\def \rr {\raise.35ex\hbox{\small $\prime$}\kern-.17em{\mbox{\large $\imath$}}}

\def \dels {\partial\kern-.6em /\kern.1em}
\def \As {{A\kern-.5em / \kern.5em}}
\def \Ds {D\kern-.7em / \kern.5em}

\def \ks {k\kern-.5em /}
\def \ls {l\kern-.5em /}

\newcommand{\ci}[1]{}

\newcommand{\ba}{\begin{eqnarray}}
\newcommand{\ea}{\end{eqnarray}}
\newcommand{\bal}{\begin{align}}
\newcommand{\eal}{\end{align}}
\newcommand{\bay}[1]{\left(\begin{array}{#1}}
\newcommand{\eay}{\end{array}\right)}

\newcommand{\hide}[1]{}

\newlist{axioms}{enumerate}{2}
\setlist[axioms,1]{label=\textbf{A\arabic{axiomsi}.}, ref=A\arabic{axiomsi}}
\setlist[axioms,2]{label=\textbf{A\arabic{axiomsi}\rlap{\myEnumCounter{axiomsii}}.},%
                   ref=A\arabic{axiomsi}\myEnumCounter{axiomsii},%
                   align=parleft,%
                   leftmargin=0em,%
                   itemsep=1.4ex,%
                   before={\stepcounter{axiomsi}}}

  \usetikzlibrary{decorations.markings}

\begin{document}
\begin{titlepage}

\begin{center}

\textbf{\LARGE
Second-Order Perturbation in\\
 Adaptive Perturbation Method
\vskip.3cm
}
\vskip .5in
{\large
Chen-Te Ma$^{a, b, c, d, e}$ \footnote{e-mail address: yefgst@gmail.com}
\\
\vskip 1mm
}
{\sl
$^a$
Asia Pacific Center for Theoretical Physics,\\
Pohang University of Science and Technology, 
Pohang 37673, 
Gyeongsangbuk-do, South Korea;
\\
$^b$
Guangdong Provincial Key Laboratory of Nuclear Science,\\
 Institute of Quantum Matter,
South China Normal University, Guangzhou 510006, Guangdong, China.
\\
$^c$
School of Physics and Telecommunication Engineering,\\ 
South China Normal University, Guangzhou 510006, Guangdong, China.
\\
$^d$
Guangdong-Hong Kong Joint Laboratory of Quantum Matter,\\
 Southern Nuclear Science Computing Center, 
South China Normal University, 
Guangzhou 510006, Guangdong, China.
\\
$^e$
The Laboratory for Quantum Gravity and Strings,\\
 Department of Mathematics and Applied Mathematics,\\
University of Cape Town, Private Bag, Rondebosch 7700, South Africa.
}
\\
\vskip 1mm
\vspace{40pt}
\end{center}
\begin{abstract}
The perturbation method is an approximation scheme with a solvable leading order. 
The standard way is to choose a non-interacting sector for the leading order. 
The adaptive perturbation method improves the solvable part by using all diagonal elements for a Fock state.
We consider the harmonic oscillator with the interacting term, $\lambda_1x^4/6+\lambda_2x^6/120$, where $\lambda_1$ and $\lambda_2$ are coupling constants, and $x$ is the position operator. 
The spectrum shows a quantitative result from the second-order, less than 1 percent error, compared to a numerical solution when turning off the $\lambda_2$. 
When we turn on the $\lambda_2$, more deviation occurs, but the error is still less than 2 percent.  
We show a quantitative result beyond a weak-coupling region. 
Our study should provide interest in the holographic principle and strongly coupled boundary theory. 
\end{abstract}
\end{titlepage}

\section{Introduction}
\label{sec:1}
\noindent
It is hard to solve quantum systems without requiring high symmetry, like a hydrogen atom and a harmonic oscillator \cite{Ma:2018efs}. 
Therefore, people used the above solvable systems to generate a more complicated system by perturbing it. 
In Quantum Field Theory (QFT), people often choose a non-interacting theory for solvable systems. 
The perturbation series cannot explore any physical significance for a large coupling constant \cite{Wilson:1971dc}. 
Hence the perturbation technology \cite{Hioe:1978jj} only probes physics in a weak-coupling region. 
However, the perturbation series is not convergent. 
This series is asymptotic convergence. 
The Borel summation solved the problem in some models but needs a complicated resummation. 
We will adopt the {\it adaptive perturbation method} \cite{Weinstein:2005kw,Weinstein:2005kx}. 
The method is simple enough to provide an {\it analytical solution} to the spectrum. 
In this letter, our calculation is only up to the {\it second-order}. 
The analytical solution is enough to show a {\it quantitative} comparison to numerical solutions in different values of coupling constants beyond a weak-coupling perturbation.   
\\

\noindent
The adaptive perturbation method chooses the diagonal elements of a Fock space for the solvable part \cite{Weinstein:2005kw}. 
This perturbation includes the interacting information at the leading order \cite{Weinstein:2005kw}. 
The unperturbed state is always a Fock state \cite{Weinstein:2005kw}. 
In the old calculation of QFT, people used the eigenstate of non-interacting theory.
The state is not a Fock state for a loss of mass term and spatial derivative terms. 
Another difference is a variable $\gamma$. 
This variable transforms the position operator by a {\it scaling factor}. 
The commutation relation between the momentum and position operators does not change \cite{Weinstein:2005kw}. 
Fixing $\gamma$ is through a minimized expectation value of Hamiltonian \cite{Weinstein:2005kw}. 
The variation method can give a proper estimation for a ground state. 
However, the leading-order result shows that the estimation is good enough for excited states in all tested coupling constants \cite{Curcio:2018,Ma:2019pxd}. 
The variation method should give a proper saddle point. 
The perturbation parameter is not a coupling constant when a mass term loses \cite{Ma:2019pxd}. 
This result shows that such a perturbation is still valid for a large coupling constant. 
The central question that we would like to address in this letter is: {\it How practical for the adaptive perturbation method?} 
\\

\noindent
We demonstrate an analytical study from the second-order perturbation. 
If this method is practical, the solution should be close to the numerical solutions. 
In the leading order, we fix the variable by minimizing the expectation value of the Hamiltonian. 
For the second-order calculation, the fixing has ambiguities. 
We can choose the same value for $\gamma$ as the leading order or fix the $\gamma$ by minimizing the eigenenergy with a second-order correction. 
However, the exact result should not depend on a choice of $\gamma$. 
The choice of $\gamma$ should rely on the convenience of the calculation and how practical. 
However, the calculation is {\it simple} and {\it practical} if one chooses the value of $\gamma$ as in the leading-order. 
\\

\noindent
In this letter, we consider a harmonic oscillator model with the potential $\lambda_1 x^4/6+\lambda_2 x^6/120$. 
The $\lambda_1$ and $\lambda_2$ are coupling constants, and $x$ is the position operator. 
We can obtain the saddle points {\it analytically} in such a model. 
Hence we can provide an {\it analytical solution} to the second-order perturbation. 
Usually, the perturbation study in a strong-coupling region is complicated, and it is hard to have an analytical solution. 
We compare the analytical solution to the numerical solutions for different values of the coupling constants. 
The {\it maximum} deviation is around 2$\%$. 
Because our calculation is only up to the second-order, the adaptive perturbation method should be practical. 
The study of strongly coupled field theory loses a general and practical method. 
Therefore, the concrete result usually only stays in the weak-coupling region. 
It is hard to directly compute the strongly coupled physics (like in the low-energy duality web and holographic principle). 
Our practical method should provide the first-principle study in the strongly-coupled field theory (helpful for studying the renormalization group flow and strongly coupled boundary theory or the semi-classical black hole in the context of the AdS/CFT correspondence). 

\section{Adaptive Perturbation Method}
\label{sec:2}
\noindent
We show a procedure of the adaptive perturbation method from the Hamiltonian 
\bea
H=\frac{p^2}{2}+\omega^2\frac{x^2}{2}+\lambda_1 \frac{x^4}{6}+\lambda_2\frac{x^6}{120}.
\eea
The $p$ is the momentum operator, and $\omega$ is the frequency. 
The commutation relation between the $x$ and $p$ is $\lbrack p, x\rbrack=-i$. 
We rewrite the $x$ and $p$ in terms of the creation operator $A_{\gamma}^{\dagger}$ and annihilation operator $A_{\gamma}$ \cite{Weinstein:2005kw}: 
\bea
x=\frac{1}{\sqrt{2\gamma}}(A_{\gamma}^{\dagger}+A_{\gamma}); \qquad p=i\sqrt{\frac{\gamma}{2}}(A_{\gamma}^{\dagger}-A_{\gamma}).
\eea 
To satisfy the commutation relation between $x$ and $p$, the creation and annihilation operators need to satisfy \cite{Weinstein:2005kw}
\bea
\lbrack A_{\gamma}, A_{\gamma}^{\dagger}\rbrack=1.
\eea 
The $\gamma$ is a scaling factor of $x$, and it does not modify $\lbrack p, x\rbrack=-i$. 
The creation and annihilation operators depend on the $\gamma$. 
Hence the vacuum state also depends on the $\gamma$, $A_{\gamma}|0_{\gamma}\rangle=0$ \cite{Weinstein:2005kw}. 
The number operator is defined similarly as the following \cite{Weinstein:2005kw}:
 \bea
 N_{\gamma}|n_{\gamma}\rangle\equiv A_{\gamma}^{\dagger}A_{\gamma}|n_{\gamma}\rangle=n|n_{\gamma}\rangle.
 \eea 
\\

\noindent
We decompose the Hamiltonian to a solvable and perturbation part \cite{Weinstein:2005kw}. 
The solvable part $H_0(\gamma)$ is the diagonal elements for a Fock space \cite{Weinstein:2005kw}. 
Other terms in the Hamiltonian are the perturbation part $V(\gamma)$ \cite{Weinstein:2005kw}. 
Hence the solvable part $H_0(\gamma)$ in $H$ contains the coupling constants. 
When applying to the time-independent perturbation in a single-particle system, the formula of the eigenenergy is  
\bea
E_n=
E_n^{(0)}+\sum_{k\neq n}
\frac{|\langle k^{(0)}|V|n^{(0)}\rangle|^2}{E_n^{(0)}-E_{k,n}^{(0)}}+\cdots.
\eea
The $E_n^{0}$ is the $n$-th unperturbed eigenenergy. 
The $|n^{(0)}\rangle$ is the $n$-th unperturbed eigenstate. 
Because $V$ is the non-diagonal elements of the Fock space, the first-order term $\langle n^{(0)}|V|n^{(0)}\rangle$ vanishes.  
We determine the unfixed variable $\gamma$ by minimizing the expectation value of the Hamiltonian $E_n(\gamma)_{\mathrm{min}}$ or $E_n^{(0)}$. 
The $E^{(0)}_{k, n}$ is the $k$-th unperturbed eigenenergy with the $n$-th unperturbed eigenstate's $\gamma$. 
Hence we determine the value of $\gamma$ by the leading order or the solvable part and continuously use the same value in the higher-order perturbation terms. 
Because each unperturbed state gives a different value of $\gamma$, using the same value in the second-order is strange. 
However, we will show that the perturbation solution can compare to the numerical solution in all tested regions.
\\

\noindent
The unperturbed spectrum is \cite{Ma:2019pxd}
\bea
&&E_n(\gamma)_{\mathrm{min}}
\nn\\
&=&\frac{\gamma}{4}(2n+1)+\frac{\omega^2}{4\gamma}(2n+1)
\nn\\
&&
+\frac{\lambda_1}{4\gamma^2}\bigg(n^2+n+\frac{1}{2}\bigg)
+\frac{\lambda_2}{4\gamma^3}\bigg(\frac{1}{12}n^3+\frac{29}{240}n^2+\frac{1}{6}n+\frac{1}{16}\bigg).
\nn\\
\eea
The expectation value of the Hamiltonian is minimal when the $\gamma$ is positive and satisfies \cite{Ma:2019pxd}
\bea
\gamma^4-\omega^2\gamma^2-\lambda_1\frac{2n^2+2n+1}{2n+1}\gamma
-\frac{\lambda_2}{80}\frac{20n^3+29n^2+40n+15}{2n+1}
=0.
\nn\\
\eea
Our direct second-order calculation gives the spectrum
\bea
&&E_n(\gamma)_2
\nn\\
&=&\frac{\gamma}{4}(2n+1)+\frac{\omega^2}{4\gamma}(2n+1)+\frac{\lambda_1}{4\gamma^2}\bigg(n^2+n+\frac{1}{2}\bigg)
+\frac{\lambda_2}{4\gamma^3}\bigg(\frac{1}{12}n^3+\frac{29}{240}n^2+\frac{1}{6}n+\frac{1}{16}\bigg)
\nn\\
&&
+\frac{\lambda_2^2}{921600\gamma^6}
\frac{(n+1)(n+2)(n+3)(n+4)(n+5)(n+6)}
{-3\gamma
-\frac{3\omega^2}{\gamma}-\frac{3\lambda_1}{2\gamma^2}(2n+7)+\frac{\lambda_2}{4\gamma^3}\bigg(-\frac{3}{2}(n^2+6n+12)-\frac{29}{20}(n+3)-1\bigg)}
\nn\\
&&+
\frac{(n+1)(n+2)(n+3)(n+4)\bigg(\frac{\lambda_2}{320\gamma^3}(2n+5)+\frac{\lambda_1}{24\gamma^2}\bigg)^2}
{-2\gamma
-\frac{2\omega^2}{\gamma}-\frac{\lambda_1}{\gamma^2}(2n+5)-\frac{\lambda_2}{4\gamma^3}\bigg\lbrack\bigg(n^2+4n+\frac{16}{3}\bigg)+\frac{29}{30}(n+2)+\frac{2}{3}\bigg\rbrack}
\nn\\
&&+
\frac{(n+1)(n+2)\bigg(-\frac{\gamma}{4}+\frac{\omega^2}{4\gamma}+\frac{\lambda_1}{12\gamma^2}(2n+3)+\frac{\lambda_2}{64\gamma^3}(n^2+3n+3)\bigg)^2}
{-\gamma
-\frac{\omega^2}{\gamma}-\frac{\lambda_1}{2\gamma^2}(2n+3)-\frac{\lambda_2}{4\gamma^3}\bigg\lbrack\bigg(\frac{1}{2}n^2+n+\frac{2}{3}\bigg)+\frac{29}{60}(n+1)+\frac{1}{3}\bigg\rbrack}
\nn\\
&&+
\frac{(n-1)n\bigg(-\frac{\gamma}{4}+\frac{\omega^2}{4\gamma}+\frac{\lambda_1}{12\gamma^2}(2n-1)+\frac{\lambda_2}{64\gamma^3}(n^2-n+1)\bigg)^2}
{\gamma+\frac{\omega^2}{\gamma}+\frac{\lambda_1}{2\gamma^2}(2n-1)+\frac{\lambda_2}{4\gamma^3}\bigg\lbrack\bigg(\frac{1}{2}n^2-n+\frac{2}{3}\bigg)+\frac{29}{60}(n-1)+\frac{1}{3}\bigg\rbrack}
\nn\\
&&+
\frac{(n-3)(n-2)(n-1)n\bigg(\frac{\lambda_1}{24\gamma^2}+\frac{\lambda_2}{320\gamma^3}(2n-3)\bigg)^2}
{2\gamma+\frac{2\omega^2}{\gamma}+\frac{\lambda_1}{\gamma^2}(2n-3)+\frac{\lambda_2}{4\gamma^3}\bigg\lbrack\bigg(n^2-4n+\frac{16}{3}\bigg)+\frac{29}{30}(n-2)+\frac{2}{3}\bigg\rbrack}
\nn\\
&&+
\frac{\lambda_2^2}{921600\gamma^6}
\frac{(n-5)(n-4)(n-3)(n-2)(n-1)n}
{3\gamma+\frac{3\omega^2}{\gamma}+\frac{3\lambda_1}{2\gamma^2}(2n-5)+\frac{\lambda_2}{4\gamma^3}\bigg(\frac{3}{2}(n^2-6n+12)+\frac{29}{20}(n-3)+1\bigg)}.
\nn\\
\eea
\\

\noindent
When one chooses: 
\bea
\omega=\lambda_2=0,
\eea 
each perturbation term is at the same order of the coupling constant $\lambda_1^{1/3}$ \cite{Ma:2019pxd}. 
This order matches with an exact result. Because we can do the transformations, 
\bea
x\rightarrow \frac{x}{\lambda_1^{\frac{1}{6}}}; \qquad p\rightarrow\lambda_1^{\frac{1}{6}}p,
\eea
 to show $H\propto\lambda_1^{1/3}$. The case: 
\bea
\omega=\lambda_1=0
\eea
 is also similar. 
 Hence the perturbation parameter should not be a coupling constant \cite{Ma:2019pxd}.    
 
 \section{Comparison for $\lambda_2=0$}
\label{sec:3}
\noindent
We discretize the kinetic-energy term as
\bea
\frac{p^2}{2}\psi\rightarrow-\frac{\psi_{j+1}-2\psi_j+\psi_{j-1}}{2a^2},
\eea
where $\psi_j$ is the eigenfunction at the site $x_j$ for the lattice theory, and $a$ is the lattice spacing. The lattice index is labeled by $j=1, 2, \cdots, n$, where $n$ is the number of lattice points. 
The lattice system has $n+1$ lattice points with a lattice size $2L$ and the periodic boundary condition as in the following:
\bea
&&-L\le x_{j}\le L; \qquad x_{0}=-L; \qquad x_{j+1}\equiv x_{j}+a; 
\nn\\
 &&\psi_{0}\equiv \psi_{n}; \qquad  2L=na.
\eea 
 We do an exact diagonalization to obtain the eigenenergies. 
The choice of lattice size and the number of lattice points is: 
\bea
L=8;\qquad n=1024.
\eea 
\\

\noindent
We first turn off the $\lambda_2$. Our second-order perturbation solution only deviates from the numerical solution within 1$\%$ in Tables \ref{c116}, \ref{c1025}, and \ref{c316}. 
When we turn off the $\lambda_1$ and $\lambda_2$, the adaptive perturbation method gives an exact solution to the harmonic oscillator with a non-vanishing mass term \cite{Weinstein:2005kw}. 
We only use the second-order perturbation to obtain the accurate result for the interaction $x^4$. 
Hence the adaptive perturbation should be practical. 
We will do a similar test for $\lambda_2\neq 0$. 
We then define $\textbf{Deviation}\ 1$ as $\bigg(100\times\big|\big((\textbf{Numerical Solution})- E_n(\gamma)_{\mathrm{min}}\big)\big/(\textbf{Numerical Solution})\big|\bigg)\%$ and $\textbf{Deviation}\ 2$ as $\bigg(100\times\big|\big((\textbf{Numerical Solution})- E_n(\gamma)_{2}\big)\big/(\textbf{Numerical Solution})\big|\bigg)\%$ in all Tables.
\\

\begin{table}[!htb]
\centering
\caption{The comparison between the perturbation and numerical solutions for the $\omega=0$, $\lambda_1=16$, and $\lambda_2=0$.}
\begin{tabular}{ |m{1em} | m{2cm}| m{2cm}|m{2cm}|m{3cm}|m{3cm}|} 
\hline
\textbf{$n$} & \textbf{$E_n(\gamma)_{\mathrm{min}}$} & \textbf{$E_n(\gamma)_{2}$}&\textbf{Numerical Solution}&\textbf{Deviation} 1&\textbf{Deviation} 2 \\ 
\hline
$0$ & 0.944& 0.929& 0.926&1.943\%&0.323\%\\ 
\hline
$1$ & 3.361& 3.324& 3.319&1.265\%&0.15\%\\ 
\hline
$2$ & 6.496& 6.53& 6.512&0.245\%&0.276\%\\ 
\hline
$3$ & 10.11& 10.211& 10.17&0.589\%&0.403\%\\ 
\hline
$4$ & 14.098& 14.266& 14.201&0.725\%&0.457\%\\ 
\hline
$5$ & 18.398& 18.636& 18.545&0.792\%&0.49\%\\ 
\hline
$6$ & 22.97& 23.281& 23.162&0.828\%&0.513\%\\ 
\hline
$7$ & 27.785& 28.172& 28.022&0.845\%&0.535\%\\ 
\hline
\end{tabular}
\label{c116}
\end{table}

\begin{table}[!htb]
\centering
\caption{The comparison between the perturbation and numerical solutions for the $\omega=0$, $\lambda_1=0.25$, and $\lambda_2=0$.}
\begin{tabular}{ |m{1em} | m{2cm}| m{2cm}|m{2cm}|m{3cm}|m{3cm}|} 
\hline
\textbf{$n$} & \textbf{$E_n(\gamma)_{\mathrm{min}}$} & \textbf{$E_n(\gamma)_{2}$}&\textbf{Numerical Solution}&\textbf{Deviation} 1&\textbf{Deviation} 2 \\ 
\hline
$0$ & 0.236& 0.232& 0.231&2.164\%&0.432\%\\ 
\hline
$1$ & 0.84& 0.831& 0.829&1.326\%&0.241\%\\ 
\hline
$2$ & 1.624& 1.632& 1.628&0.245\%&0.245\%\\ 
\hline
$3$ & 2.527& 2.552& 2.543&0.629\%&0.353\%\\ 
\hline
$4$ & 3.524& 3.566& 3.551&0.76\%&0.422\%\\ 
\hline
$5$ & 4.599& 4.659& 4.637&0.819\%&0.474\%\\ 
\hline
$6$ & 5.742& 5.82& 5.792&0.863\%&0.483\%\\ 
\hline
$7$ & 6.946& 7.043& 7.009&0.898\%&0.485\%\\ 
\hline
\end{tabular}
\label{c1025}
\end{table}

\begin{table}[!htb]
\centering
\caption{The comparison between the perturbation and numerical solutions for the $\omega=1$, $\lambda_1=16$, and $\lambda_2=0$.}
\begin{tabular}{ |m{1em} | m{2cm}| m{2cm}|m{2cm}|m{3cm}|m{3cm}|} 
\hline
\textbf{$n$} & \textbf{$E_n(\gamma)_{\mathrm{min}}$} & \textbf{$E_n(\gamma)_{2}$}&\textbf{Numerical Solution}&\textbf{Deviation} 1&\textbf{Deviation} 2 \\ 
\hline
$0$ & 1.041& 1.029&1.026&1.461\%&0.292\%\\ 
\hline
$1$ & 3.607& 3.576&3.571&1.008\%&0.14\%\\ 
\hline
$2$ & 6.852& 6.878&6.863&0.16\%&0.218\%\\ 
\hline
$3$ & 10.56& 10.646&10.611&0.48\%&0.329\%\\ 
\hline
$4$ & 14.631& 14.78&14.723&0.624\%&0.387\%\\ 
\hline
$5$ & 19.009& 19.224&19.143&0.699\%&0.423\%\\ 
\hline
$6$ & 23.655& 23.938&23.83&0.734\%&0.453\%\\ 
\hline
$7$ & 28.539& 28.895&28.758&0.761\%&0.476\%\\ 
\hline
\end{tabular}
\label{c316}
\end{table}

\section{Comparison for $\lambda_2\neq 0$}
\label{sec:4}
\noindent
Now we turn on the interacting term $x^6$. 
The maximum deviation for the second-order perturbation becomes 1.742$\%$ in Tables \ref{c216256}, \ref{c20254}, and \ref{c416256}. 
The reason is possibly due to more transitions of the $x^6$ than the $x^4$. 
Hence suppressing the deviation to 1$\%$ is necessary from a higher-order calculation. 
However, the deviation is good enough for a qualitative study. 
Introducing a mass term reduces the deviation dramatically in a weak-coupling region by comparing Table \ref{c216256} to Table \ref{c416256}. 
We find that the case $\lambda_2\neq 0$ is similar to the study of $\lambda_2=0$. 
The comparison should show that the adaptive perturbation method is practical because the second-order is enough for a quantitative result. 
\\

\begin{table}[!htb]
\centering
\caption{The comparison between the perturbation and numerical solutions for the $\omega=0$, $\lambda_1=16$, and $\lambda_2=256$.}
\begin{tabular}{ |m{1em} | m{2cm}| m{2cm}|m{2cm}|m{3cm}|m{3cm}|} 
\hline
\textbf{$n$} & \textbf{$E_n(\gamma)_{\mathrm{min}}$} & \textbf{$E_n(\gamma)_{2}$}&\textbf{Numerical Solution}&\textbf{Deviation} 1&\textbf{Deviation} 2 \\ 
\hline
$0$ & 1.117& 1.086& 1.075&3.906\%&1.023\%\\ 
\hline
$1$ & 4.047& 3.964&3.949&2.481\%&0.379\%\\ 
\hline
$2$ & 7.993& 8.032&7.989&0.05\%&0.538\%\\ 
\hline
$3$ & 12.724& 12.939&12.831&0.833\%&0.841\%\\ 
\hline
$4$ & 18.109& 18.532&18.338&1.248\%&1.057\%\\ 
\hline
$5$ & 24.067& 24.723&24.426&1.469\%&1.215\%\\ 
\hline
$6$ & 30.54& 31.453&31.038&1.604\%&1.337\%\\ 
\hline
$7$ & 37.486& 38.67&38.13&1.688\%&1.416\%\\ 
\hline
\end{tabular}
\label{c216256}
\end{table}

\begin{table}[!htb]
\centering
\caption{The comparison between the perturbation and numerical solutions for the $\omega=0$, $\lambda_1=0.25$ and $\lambda_2=4$.}
\begin{tabular}{ |m{1em} | m{2cm}| m{2cm}|m{2cm}|m{3cm}|m{3cm}|} 
\hline
\textbf{$n$} & \textbf{$E_n(\gamma)_{\mathrm{min}}$} & \textbf{$E_n(\gamma)_{2}$}&\textbf{Numerical Solution}&\textbf{Deviation} 1&\textbf{Deviation} 2 \\ 
\hline
$0$ & 0.343& 0.331& 0.326&5.214\%&1.533\%\\ 
\hline
$1$ & 1.258& 1.225&1.218&3.284\%&0.574\%\\ 
\hline
$2$ & 2.512& 2.524&2.507&0.199\%&0.678\%\\ 
\hline
$3$ & 4.039& 4.123&4.079&0.98\%&1.07\%\\ 
\hline
$4$ & 5.795& 5.963&5.884&1.512\%&1.342\%\\ 
\hline
$5$ & 7.753& 8.015&7.894&1.786\%&1.532\%\\ 
\hline
$6$ & 9.892& 10.256&10.089&1.952\%&1.655\%\\ 
\hline
$7$ & 12.197& 12.671&12.454&2.063\%&1.742\%\\ 
\hline
\end{tabular}
\label{c20254}
\end{table}

\begin{table}[!htb]
\centering
\caption{The comparison between the perturbation and numerical solutions for the $\omega=1$, $\lambda_1=16$, and $\lambda_2=256$.}
\begin{tabular}{ |m{1em} | m{2cm}| m{2cm}|m{2cm}|m{3cm}|m{3cm}|} 
\hline
\textbf{$n$} & \textbf{$E_n(\gamma)_{\mathrm{min}}$} & \textbf{$E_n(\gamma)_{2}$}&\textbf{Numerical Solution}&\textbf{Deviation} 1&\textbf{Deviation} 2 \\ 
\hline
$0$ & 1.195& 1.168&1.159&3.106\%&0.776\%\\ 
\hline
$1$ & 4.242& 4.165&4.154&2.118\%&0.264\%\\ 
\hline
$2$ & 8.266& 8.297&8.26&0.072\%&0.447\%\\ 
\hline
$3$ & 13.059& 13.253&13.157&0.744\%&0.729\%\\ 
\hline
$4$ & 18.498& 18.889&18.712&1.143\%&0.945\%\\ 
\hline
$5$ & 24.503& 25.12&24.844&1.372\%&1.11\%\\ 
\hline
$6$ & 31.019& 31.885&31.496&1.514\%&1.235\%\\ 
\hline
$7$ & 38.005& 39.143&38.625&1.605\%&1.341\%\\ 
\hline
\end{tabular}
\label{c416256}
\end{table}
\newpage

\section{Outlook}
\label{sec:5}
\noindent
The anharmonic oscillator model is non-integrable. 
Therefore, the perturbation series to all orders cannot have an exact solution. 
However, a quantitative result appears in the second order. 
This result should imply that this method is quite efficient. 
To understand our world, we must face non-integrable models. 
Therefore, it is impossible to have an exact solution always. 
The adaptive perturbation method \cite{Weinstein:2005kw} can provide an analytical solution to each perturbed term for all coupling constants. 
Especially for a strongly coupled theory, it is hard to have an analytical study from a traditional method. 
We demonstrated that the adaptive perturbation method is simple enough to provide an analytical solution.   
\\

\noindent
Our consideration is extendable to the scalar field theory \cite{Weinstein:2005kw}. 
It is interesting to study critical points by using a renormalization group flow \cite{Wilson:1971dc}.

\section*{Acknowledgments}
\noindent
The author would like to thank Bo Feng, Gang Yang, and Ellis Ye Yuan for their discussion. 
The author also thanks Nan-Peng Ma for his encouragement.
\\

\noindent
The author acknowledges the YST Program of the APCTP; 
Post-Doctoral International Exchange Program; 
China Postdoctoral Science Foundation, Postdoctoral General Funding: Second Class (Grant No. 2019M652926); 
Foreign Young Talents Program (Grant No. QN20200230017).
\\

\noindent
The author would like to thank the Institute of Theoretical Physics at the Chinese Academy of Sciences; 
Sun Yat-Sen University. 




\end{document}